\documentclass[12pt]{article}

\usepackage{graphicx}
\begin{document}

\begin{center}
{\bf Einstein$-$Gauss$-$Bonnet gravity with nonlinear electrodynamics} \\
\vspace{5mm} S. I. Kruglov
\footnote{E-mail: serguei.krouglov@utoronto.ca}
\underline{}
\vspace{3mm}

\textit{Department of Physics, University of Toronto, \\60 St. Georges St.,
Toronto, ON M5S 1A7, Canada\\
Department of Chemical and Physical Sciences, University of Toronto,\\
3359 Mississauga Road North, Mississauga, ON L5L 1C6, Canada} \\
\vspace{5mm}
\end{center}
\begin{abstract}
We obtain an exact spherically symmetric and magnetically charged black hole solution in 4D Einstein$-$Gauss$-$Bonnet gravity coupled with nonlinear electrodynamics with the Lagrangian ${\cal L}_{NED} = -{\cal F}/(1+\sqrt{2\mid\beta{\cal F}\mid})$ (${\cal F}$ is the field strength invariant). The thermodynamics of the black hole is studied within our model. We calculate the Hawking temperature and the heat capacity of the black hole. The phase transitions occur in the point where the Hawking temperature possesses an extremum and the heat capacity diverges. It was shown that black holes are thermodynamically stable in some range of event horizon radii when the heat capacity and Hawking temperature are positive. The logarithmic correction to the Bekenstein$-$Hawking BH entropy is obtained which mimics a quantum correction. The dependence of the BH shadow radius on the model parameters is studied. We also investigate the particle emission rate of the BH. The stability of the BH is studied by analysing the quasinormal modes.

\end{abstract}

\section{Introduction}

The observational data around black holes (BHs), with the strong gravity regime, show the possibility of alternatives to the Einstein General Relativity (GR). At the same time, the heterotic string theory at the low energy limit results in models of gravity with the action including higher order curvature terms \cite{Witten}. In the paper of \cite{Glavan}, the authors proposed a new theory of gravity in four dimensions, named ``4D Einstein$-$Gauss$-$Bonnet gravity” (4D EGB), with higher order corrections. The 4D EGB gravity theory is based on equal footing with the Einstein GR.
Firstly, the spherically symmetric static BH solution for the EGB gravity was  obtained in \cite{Deser}.
The action of the 4D EGB theory includes the Einstein$-$Hilbert action plus the Gauss-Bonnet (GB) term and is the simplest case of the Lovelock theory. The Lovelock gravity is the generalization of Einstein’s GR to higher dimensions that produces covariant second-order field equations.
The 4D EGB theory gets around the Lovelock theorem \cite{Lovelock} that tells: ``when implying a 4D space-time, diffeomorphism invariance, metricity and second-order equations of motion, then GR with a cosmological constant is the unique theory of gravity". It worth noting that the GB invariant is a topological invariant in 4D and it contributes to components of equations of motion to be proportional to $D-4$. However, the authors of \cite{Glavan} shown that if re-scaling the coupling constant $\alpha$, which can be considered as the inverse of
the string tension, by $\alpha /(D-4)$ and taking the limit $D\rightarrow 4$, the theory leads to non-trivial dynamics and without singularities. In addition, the 4D EGB theory, being a classical modified gravity, is free from Ostrogradsky instability and conserves the number of degrees of freedom. The static spherically symmetric BH solutions obtained are different from the vacuum Schwarzschild BH solution and give a repulsive gravitational force at short distances. The GB term can be considered as a quantum correction to GR \cite{Deser} although the 4D EGB theory is a classical theory. Previously similar solutions were obtained in GR with a conformal anomaly \cite{Cai}, \cite{Cai1} and in gravity with quantum corrections \cite{Cognola}. Some aspects of the 4D EGB gravity theory were considered in \cite{Fernandes}- \cite{Hennigar}. The BH solutions in the 4D EGB model coupled to nonlinear electrodynamics (NED) were also studied in \cite{Martinez}-\cite{Lobo}.

It was shown in \cite{Tomozawa} that for a static spherically symmetric ansatz, in a conformally flat metric, only the GB term gives the finiteness of renormalization. The metric obtained is the rigorous solution for quantum gravity corrections and at short distances the repulsive force occurs. Note that the GB term plays an important role in 4D AdS spacetime within the AdS/CFT correspondence. A boundary term (the topological GB term) in the holographic renormalization leads to the standard thermodynamics for AdS BHs.

The 4D EGB gravity theory was debated in \cite{Tekin}-\cite{Hohmann}. In \cite{Tekin}, \cite{Hohmann} the authors note that in the action the GB term gives the contribution to field equations as $G_{\mu\nu}=(D-4)A_{\mu\nu}+W_{\mu\nu}$. The first term can be renormalized in 4D having a limit for $D \rightarrow 4$ . The second term cannot be obtained from any action and does not have a limit for $D\rightarrow 4$.
It was pointed in \cite{Tekin1} that the theory makes sense in static spherically symmetric spacetime.
The authors of \cite{Aoki}, \cite{Aoki1} also noted that the dimensional regularization of \cite{Glavan} is justified for some class of metric.
A more complete overview of the results in 4D EGB gravity were presented in recent paper \cite{Wei1}.
It was stated in \cite{Lobo1} that the solution in the framework of \cite{Aoki}, \cite{Aoki1}, in the spherically-symmetric metrics, is a solution in the scheme of dimensional regularization \cite{Glavan}. Therefore, we use here the dimension regularization considered in \cite{Glavan}.
To analyse the stability of BHs one can study the quasinormal modes which represent the response of BHs on external perturbations. The corresponding frequencies depend on the model parameters. To study the quasinormal modes we explore the WKB approximation proposed in \cite{Schutz} (see also \cite{Iyer}, \cite{Konoplya}).

Here, we obtain a BH solution in the 4D EGB model coupled to NED proposed in \cite{Kruglov1}. The advantages of the present model from other NED models \cite{Kruglov} is due to its simplicity and the obtained solution in 4D EGB gravity for the metric function is expressed through elementary functions. In addition, the stress-energy tensor in this NED model satisfies the weak and dominant energy conditions (see Appendix of the present paper). The NED model \cite{Kruglov1} was also used for studying the modified Hayward BH in \cite{Habib}.
It should be noted that NED is used in cosmology and BH physics \cite{Kruglov}-\cite{Quiros}.

The paper has a structure as follows: In Sec. 2, we obtain spherically symmetric solutions in the 4D EGB model coupled to NED. It was shown that the metric function $f(r)$ in the limit $r\rightarrow 0$ leads to the reasonable value $f(0)=1$. The metric function can have one or two BH horizons or no horizons when the BH does not exists depending on the model parameters. The Hawking temperature and the heat capacity are calculated and the stability and phase transitions are studied in Sec. 3. It was demonstrated that the BHs are thermodynamically stable in some range of event horizon radii. We show that the entropy in our model contains the logarithmic correction to the Bekenstein$-$Hawking
area low which mimics the quantum correction. The BH shadow is studied in Sec. 4. In Sec. 5 we investigate the particle emission rate of the BH. The quasinormal modes were studied in Sec. 6. Section 7 is a conclusion. In Appendix we study energy conditions.

\section{EGB model with NED}

The EGB gravity in D-dimensions coupled to NED is described by the action
\begin{equation}
I=\int d^Dx\sqrt{-g}\left[\frac{1}{16\pi G}\left(R+ \frac{\alpha}{D-4}{\cal L}_{GB}\right)+{\cal L}_{NED}\right],
\label{1}
\end{equation}
where $\alpha$ has the dimension of (length)$^2$ and the Lagrangian of NED, proposed in \cite{Kruglov1}
\footnote{For a convenience we substituted $\beta$ in \cite{Kruglov1} by $2\beta$ and to consider pure electric field we use the absolute value
$\mid\beta{\cal F}\mid$. In this model the finite electric field in the origin of charges holds.} is given by
\begin{equation}
{\cal L}_{NED} = -\frac{{\cal F}}{1+\sqrt{2\mid\beta{\cal F}\mid}},
 \label{2}
\end{equation}
where the parameter $\beta$ ($\beta\geq 0$) possesses the dimension of (length)$^4$, ${\cal F}=(1/4)F_{\mu\nu}F^{\mu\nu}=(B^2-E^2)/2$, $F_{\mu\nu}=\partial_\mu A_\nu-\partial_\nu A_\mu$ is the field strength tensor.
It is worth noting that $\mid\beta{\cal F}\mid$ violates the analytical properties of the Lagrangian (2) at ${\cal F}=0$. Thus, the derivatives of ${\cal L}_{NED}$ possess the discontinuity. Also, it could lead to a problem to describe the propagation of electromagnetic waves in vacuum. Similar problems exist in the model \cite{Bronnikov}. But for our case here ${\cal F}>0$, and ${\cal L}_{NED}$ is good-defined.
The GB Lagrangian is
\begin{equation}
{\cal L}_{GB}=R^{\mu\nu\alpha\beta}R_{\mu\nu\alpha\beta}-4R^{\mu\nu}R_{\mu\nu}+R^2.
\label{3}
\end{equation}
By varying of action (1) with respect to the metric one obtains the field equations
\begin{equation}
R_{\mu\nu}-\frac{1}{2}g_{\mu\nu}R+\frac{\alpha}{D-4}H_{\mu\nu}=-8\pi GT_{\mu\nu},
\label{4}
\end{equation}
where
\begin{equation}
H_{\mu\nu}=2\left(RR_{\mu\nu}-2R_{\mu\alpha}R^\alpha_{~\nu}-2R_{\mu\alpha\nu\beta}R^{\alpha\beta}-
R_{\mu\alpha\beta\gamma}R^{\alpha\beta\gamma}_{~~~\nu}\right)-\frac{1}{2}{\cal L}_{GB}g_{\mu\nu}.
\label{5}
\end{equation}
The symmetrical stress-energy tensor of our NED \cite{Kruglov1} is given by
\begin{equation}
T_{\mu\nu}=-\frac{(2+\sqrt{2\beta{\cal F}})F_\mu^{~\alpha}F_{\nu\alpha}}{2(1+\sqrt{2\beta{\cal F}})^{2}}
-g_{\mu\nu}{\cal L}_{NED}.
\label{6}
\end{equation}
As was proven in \cite{Bronnikov}, \cite{Bronnikov1} GR coupled to NED, possessing the Maxwell asymptotic as ${\cal F}\rightarrow 0$, does not admit a static, spherically symmetric solution with a regular centre and a nonzero electric charge. Therefore, we consider here a magnetic BH. From Eq. (6) the magnetic energy density  is \cite{Kruglov1}
\begin{equation}
\rho=T_t^{~t}=\frac{B^2}{2(\sqrt{\beta} B+1)}=\frac{q_m^2}{2r^2(r^2+ q_m\sqrt{\beta})},
\label{7}
\end{equation}
where $q_m$ is a magnetic charge, ${\cal F}=q_m^2/(2r^4)$.
At the limit $D \rightarrow 4$ the $tt$ component of the field equation (4) gives
\begin{equation}
r(2\alpha f(r)-r^2-2\alpha)f'(r)-(r^2+\alpha f(r)-2\alpha)f(r)+r^2-\alpha=\frac{q_m^2r^2G}{r^2+\sqrt{\beta }q_m}.
\label{8}
\end{equation}
 The static and spherically symmetric metric reads
\begin{equation}
ds^2=-f(r)dt^2+\frac{1}{f(r)}dr^2+r^2(d\vartheta^2+\sin^2\vartheta d\phi^2),
\label{9}
\end{equation}
with the metric function $f(r)$.
The solution to Eq. (8) is given by
\begin{equation}
f(r)=1+\frac{r^2}{2\alpha}\left(1-\sqrt{1+\frac{8M\alpha G}{r^3}+\frac{4\alpha q_m^{3/2}G}{\beta^{1/4}r^3}\arctan\left(\frac{r}{\sqrt[4]{\beta q_m^2}}\right)}\right).
\label{10}
\end{equation}
The solution (10) in terms of the dimensionless variable $x=r/\sqrt[4]{\beta q_m^2}$ becomes
\begin{equation}
f(x)=1+cx^2-c\sqrt{x^4+x(a+b\arctan(x))},
\label{11}
\end{equation}
where
\begin{equation}
a=\frac{8M\alpha G}{\beta^{3/4}q_m^{3/2}},~~~b=\frac{4\alpha G}{\beta},~~~c=\frac{\sqrt{\beta}q_m}{2\alpha}.
\label{12}
\end{equation}
Here, $8M\alpha G$ is the constant of integration with $M$ being the Schwarzschild BH mass. There are two branches with signs plus and minus before the square root in Eqs. (10) and (11). We use the sign minus to have the stable BH and without ghosts (in this regards see \cite{Deser}).
The asymptotic of the metric function $f(r)$ (10) as $r\rightarrow 0$ and $r\rightarrow \infty$ are given by
\begin{equation}
f(r)=1+\sqrt{\frac{2MG}{\alpha}}\sqrt{r}+\frac{r^2}{2\alpha}+{\cal O}(r^3)~~~~r\rightarrow 0,
\label{13}
\end{equation}
\begin{equation}
f(r)=1-\frac{2mG}{r}+\frac{Gq_m^2}{r^2}+{\cal O}(r^{-3})~~~~r\rightarrow \infty,
\label{14}
\end{equation}
where
\[
m=M+\frac{\pi q_m^{3/2}}{4\beta^{1/4}}\equiv M+m_m,
\]
and $m$ is the total mass of the BH and $m_m=\int_0^\infty r^2\rho dr=\pi q_m^{3/2}/(4\beta^{1/4})$ is the electromagnetic mass. In accordance with Eq. (13) we have the regular BH because $f(r)\rightarrow 1$ as $r\rightarrow 0$. Equation (14) shows the Reissner$-$Nordstr\"{o}m behaviour of the charged BH at infinity.

It should be noted that the limit $\beta\rightarrow 0$ has to be made in Eq. (8) before the integration. In this case the solution to Eq. (8) at $\beta=0$ becomes \cite{Fernandes}
\begin{equation}
f(r)=1+\frac{r^2}{2\alpha}\left(1-\sqrt{1+\frac{8M\alpha G}{r^3}-\frac{4\alpha q_m^2G}{r^4}}\right).
\label{15}
\end{equation}
Note that the limit $r\rightarrow 0$ in Eq. (15) gives the non-physical complex value of the metric function $f(r)$, but the limit $r\rightarrow 0$ in Eqs. (10) and (11) leads to the satisfactory value $f(0)=1$. The plot of the function (11) is given in Fig. 1.
\begin{figure}[h]
\includegraphics[height=4.0in,width=4.0in]{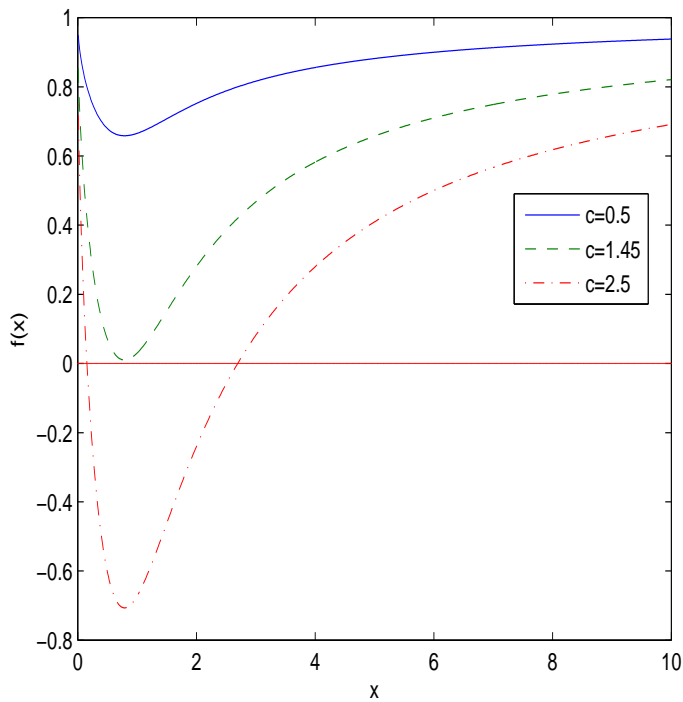}
\caption{\label{fig.1}The plot of the function $f(x)$ for $a=b=1$.}
\end{figure}
According to Fig. 1 there can be two horizons, one extreme horizon or not horizons when the BH does not exist.

\section{The BH stability and thermodynamics}

To study the BH thermodynamics and the thermal stability we calculate the Hawking temperature which is given by
\begin{equation}
T_H(r_+)=\frac{1}{4\pi}f'(r)\mid_{r=r_+},
\label{16}
\end{equation}
where $r_+$ is the event horizon radius obeying the equation $f(r_+)=0$. Making use of Eq. (11) ($x=r/\sqrt[4]{\beta q_m^2}$) we obtain the Hawking temperature
\begin{equation}
 T_H(x_+)=\frac{1}{4\pi \sqrt[4]{\beta q_m^2}}\left(\frac{(2cx_+^2-1)(1+x_+^2)-bc^2x_+^2}{2x_+(1+x_+^2)(1+cx_+^2)}\right),
 \label{17}
\end{equation}
where we replaced the parameter $a$ in Eq. (16) from equation $f(x_+)=0$. The plot of the dimensionless function $T_H(x_+)\sqrt[4]{\beta q_m^2}$ is depicted in Fig. 2.
\begin{figure}[h]
\includegraphics[height=4.0in,width=4.0in]{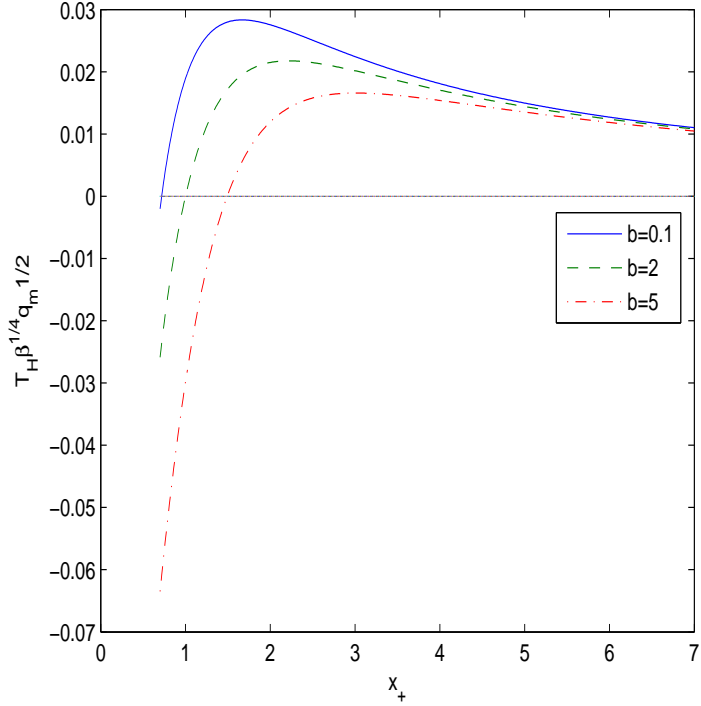}
\caption{\label{fig.2}The plot of the function $T_H(x_+)\sqrt[4]{\beta q_m^2}$ at $c=1$, $b=0.1, 2, 5$.}
\end{figure}
In accordance with Fig. 2 the Hawking temperature is positive in some interval of $x_+$ and the BH does not exist if the Hawking temperature is negative. We obtain the BH gravitational mass by solving the equation $f(x_+)=0$,
\begin{equation}
M(x_+)=\frac{\beta^{3/4}q_m^{3/2}}{8\alpha G}\left(\frac{1+2cx_+^2}{c^2x_+}-b\arctan(x_+)\right).
 \label{18}
\end{equation}
Taking into account the first law of BH thermodynamics
\begin{equation}
dM(x_+)=T_H(x_+)dS+\Phi dq,
\label{19}
\end{equation}
 the entropy at the constant charge becomes
\begin{equation}
S=\int \frac{dM(x_+)}{T_H(x_+)}=\int \frac{1}{T_H(x_+)}\frac{\partial M(x_+)}{\partial x_+}dx_+.
\label{20}
\end{equation}
With the help of Eq. (20) we find the expression for the heat capacity
\begin{equation}
C_q(x_+)=T_H\left(\frac{\partial S}{\partial T_H}\right)_q=\frac{\partial M(x_+)}{\partial T_H(x_+)}=\frac{\partial M(x_+)/\partial x_+}{\partial T_H(x_+)/\partial x_+}.
\label{21}
\end{equation}
Making use of Eqs. (17) and (18) one obtains
\begin{equation}
\frac{\partial M(x_+)}{\partial x_+}=\frac{\beta^{3/4}q_m^{3/2}}{8\alpha G}\left(\frac{2cx_+^2-1}{c^2x_+^2}-\frac{b}{1+x_+^2}\right),
\label{22}
\end{equation}
\begin{equation}
\frac{\partial T_H(x_+)}{\partial x_+}=\frac{1}{4\pi \sqrt[4]{\beta q_m^2}}\biggl(\frac{5cx_+^2-2c^2x_+^4+1}{2x_+^2(1+cx_+^2)^2}
+\frac{bc^2(3cx_+^4+(c+1)x_+^2-1)}{2(1+x_+^2)^2(1+cx_+^2)^2}\biggr).
\label{23}
\end{equation}
According to Eq. (21) the heat capacity possesses a singularity when the Hawking temperature has an extremum ($\partial T_H(x_+)/\partial x_+=0$). By virtue of Eqs. (21), (22) and (23) we depict the heat capacity versus the variable $x_+$ in Fig. 3.
\begin{figure}[h]
\includegraphics[height=4.0in,width=4.0in]{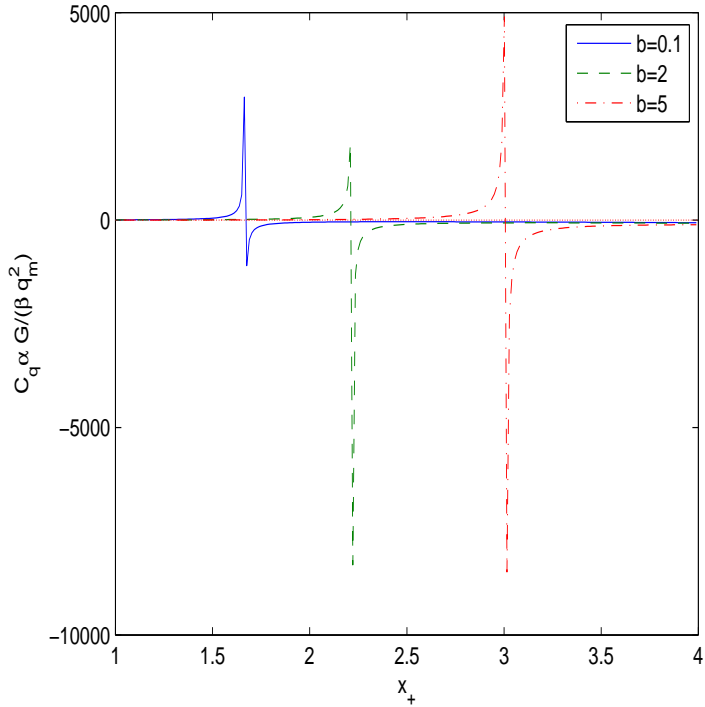}
\caption{\label{fig.3}The plot of the function $C_q(x_+)\alpha G/(\beta q_m^2)$ at $c=1$, $b=0.1, 2, 5$. }
\end{figure}
In accordance with Fig. 3 the BH is locally stable in some range of the horizon radii where the heat capacity is positive. The singularity in the heat capacity takes place in the points where the Hawking temperature has the extremum indicating on the second-order phase transition. According to Eq. (21) at some point, $x_1$, the specific heat as well as Hawking temperature become zero indicating a first-order phase transition.
In this point $x_+=x_1$ we have the BH remnant, so that the BH mass is not zero but the Hawking temperature and heat capacity vanish.
From Eq. (22) (as well as from Eq. (17)) we obtain that  $x_1$ obeys the biquadratic equation $2cx_1^4+(2c-bc^2-1)x_1^2-1=0$. The solution $x_2$  to equation $\partial T_H/\partial x_+=0$ gives the discontinuity point of the heat capacity. As a result, the BHs are locally stable at the range $x_2>x_+>x_1$. When $x_+>x_2$ the BH evaporates and is unstable. At the point $x_+=x_2$ the evaporation stops and at the range $x_2>x_+>x_1$ the BH becomes stable. At the point $x_+=x_1$ the BH remnant is formed.

Making use of Eqs. (17), (20) and (22) we obtain the entropy
\begin{equation}
S=\frac{4\pi \alpha}{G}\int\frac{1+cx_+^2}{x_+}dx_+=
\frac{\pi r_+^2}{G}+\frac{4\pi\alpha}{G}\ln\left(\frac{r_+}{\sqrt[4]{\beta q_m^2}}\right).
\label{24}
\end{equation}
The constant of integration in Eq. (24) was chosen zero. Equation (24) shows that the entropy in our model besides the Bekenstein$-$Hawking area low includes the logarithmic correction. It should be noted that similar logarithmic term is also present in GR with a conformal
anomaly, loop quantum gravity and string theory \cite{Cai}-\cite{Cognola}. Thus, the logarithmic correction obtained mimics the quantum corrections which are due to the GB Lagrangian.
When the coupling constant of GB term $\alpha$ vanishes we come to the Bekenstein$-$Hawking entropy.
For the massive BHs with the large event horizon radius $x_+$ the leading term to the entropy is the Bekenstein$-$Hawking area law. But for small $x_+$ (for the light BHs) the quantum corrections take place and the logarithmic term is important.
It is worth noting that at some values of parameters $\alpha$, $\beta$, $q_m$  and $r_+$ the entropy is zero. The event horizon radius, when the entropy vanishes, is the solution of equation ($S=0$) $r_+^4\exp(r_+^2/\alpha)=\beta q_m^2$ which is $r_+=\sqrt{2\alpha W_0(\sqrt{\beta} q_m/(2\alpha))}$, where $W_0(x)$ is the Lambert function. It is interesting that at $r_+<r_0$ ($r_0=\sqrt{2\alpha W_0(\sqrt{\beta} q_m/(2\alpha))}$) the entropy becomes negative which can indicate a new type of instability. Probably, for such values of $r_0$ the BH does not exist. It is worth noting that the negative entropy of BHs was discussed in \cite{Odintsov3}.

\section{The black hole shadow}

The BH shadow is due to the gravitational light lensing. The first image of the super-massive M87* BH was received by the Event Horizon Telescope collaboration  \cite{Event1}. The shadow of a neutral Schwarzschild BH was investigated in \cite{Synge}. We will study the magnetically charged BH shadow in the framework of 4D EGB coupled to NED described by Eq. (2). The Hamilton$-$Jacobi method for the description of the photon motion in the static spherically symmetric spacetime will be used. Let us consider the photons moving in the equatorial plane with $\vartheta=\pi/2$.
With the help of the Hamilton$-$Jacobi approach for null curves, the photon motion is given by the equation \cite{Zhang1}
\begin{equation}
H=\frac{1}{2}g^{\mu\nu}p_\mu p_\nu=\frac{1}{2}\left(\frac{L^2}{r^2}-\frac{E^2}{f(r)}+
\frac{\dot{r}^2}{f(r)}\right)=0,
\label{25}
\end{equation}
where $p_\mu$ being the photon momenta, $\dot{r}=\partial H/\partial p_r$, $E=-p_t$ is the energy and $L=p_\phi$ is the angular momentum of the photon and they are the constants of motion. From Eq, (25) one can obtain
\begin{equation}
V+\dot{r}^2=0, ~~~V=f(r)\left(\frac{L^2}{r^2}-\frac{E^2}{f(r)}\right).
\label{26}
\end{equation}
The null geodesics of the equatorial circular motion satisfy $\dot{r}=0$ and $\ddot{r}=0$ where the overdot denotes
a derivative with respect to the affine parameter. These equations lead to $V(r)=0$ and $V'(r)=0$ with the stability of a circular null geodesic at $V''(r)>0$ or $V''(r)<0$ for an unstable motion.
Thus, the equation $V(r_p)=V'(r)_{|r=r_p}=0$, corresponds to the circular orbit radius $r_p$ of the photon. Making use of these equations, we find
\begin{equation}
\xi\equiv\frac{L}{E}=\frac{r_p}{\sqrt{f(r_p)}},~~~f'(r_p)r_p-2f(r_p)=0,
\label{27}
\end{equation}
where $\xi$ denotes the impact parameter.
We use the numerical method to solve Eq. (27) for obtaining the radius of the photon sphere.
The BH shadow radius $r_s$ seen by a static observer in the position $r_0$ is given by \cite{Zhang1}
\begin{equation}
r_s=r_p\sqrt{\frac{f(r_0)}{f(r_p)}}.
\label{28}
\end{equation}
As $r_0\rightarrow \infty$, for a distant observer, one has to put $f(r_0)=1$ in Eq. (28) and, as a result, the shadow radius is equal to $r_s=\xi$. The biggest root of the equation $f(r_h)=0$  defines the event horizon radius $x_+$. Making use of Eq. (11) and $f(r_h)=0$, one finds the parameters $a$, $b$ and $c$ depending on the horizon radii
\begin{equation}
a=\frac{1+2cx_h^2-c^2x_hb\arctan(x_h)}{c^2x_h},~~~b=\frac{1+2cx_h^2-c^2x_ha}{c^2x_h\arctan(x_h)},
\label{29}
\end{equation}
\begin{equation}
c=\frac{x_h^2+\sqrt{x_h^4+x_h(a+b\arctan(x_h))}}{x_h(a+b\arctan(x_h))},
\label{30}
\end{equation}
with $x_h=r_h/\sqrt[4]{\beta q_m^2}$.
The plots of the functions (29) at $b=0.1,1,2$, $c=1$ and at $a=0.1,1,2$, $c=1$ and (30) at $a=2, b=2,3,4$ are given in Fig. 4.
\begin{figure}[h]
\includegraphics[height=4.0in,width=4.0in]{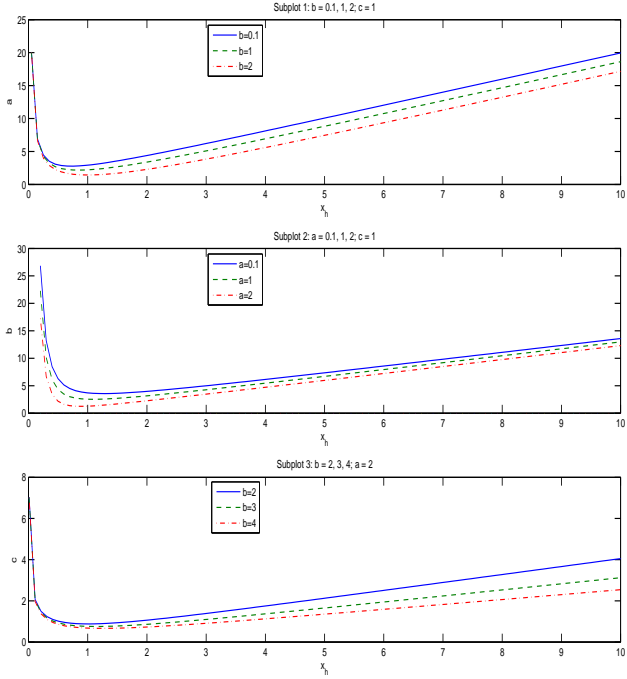}
\caption{\label{fig.4}The plots of the functions $a(x_h)$, $b(x_h)$, $c(x_h)$.}
\end{figure}
According to Fig. 4, Subplot l, if the parameter $a$ increases (at $c=1$, and fixed $b$), the event horizon radius $x_+$  increases. In accordance with Fig. 4, Subplot 2, when the parameter $b$ increases (at fixed $a$, $c=1$), the event horizon radius also increases. Figure Fig. 4, Subplot 3, shows that increasing $c$ (at fixed $a=2$, $b$) the event horizon radius  increases.
The shadow radius is formed by the null geodesics around the unstable circular photon orbit. The inner region of shadow is formed by the null geodesics that are captured by BH and they are located inside the event horizon.
The photon sphere radii ($x_p$), the event horizon radii ($x_+$), and the shadow radii ($x_s$) for some parameters  $a$, $b$, and $c$, found from Eqs. (27) and (28), are presented in Tables 1 and 2 (in terms of dimensionless variables). The null geodesics radii $x_p$ in Tables 1 and 2 correspond to maximum of the potential $V(r)$ ($V''\leq 0$) and they belong to unstable orbits. There are also solutions to Eq. (27) with $V''\geq 0$ and the corresponding radii  are less then event horizon radii and such geodesics are inside the event horizon (see Fig. 5).
\begin{figure}[h]
\includegraphics[height=4.0in,width=4.0in]{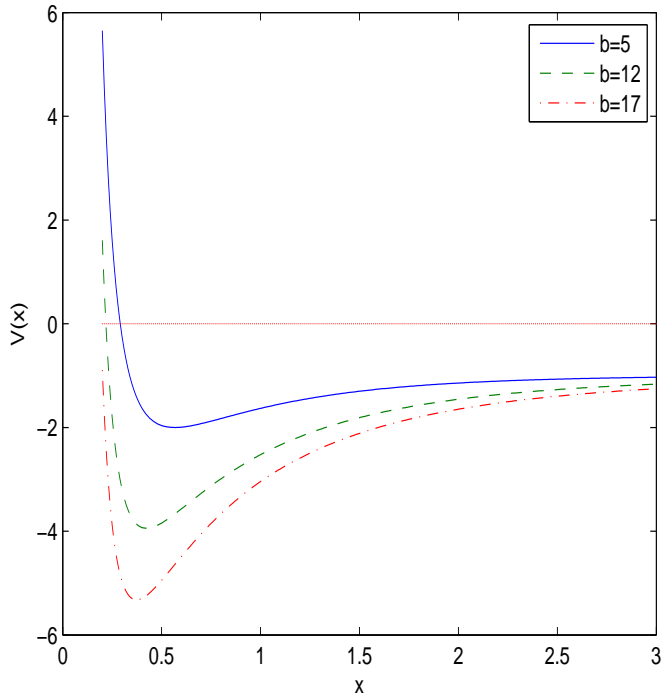}
\caption{\label{fig.5}The plot of the function $V(x)$ (at $L/\sqrt[4]{\beta q_m^2}=E=1$ for $a=2$, $b=5,12,17$, $c=1$.)}
\end{figure}
\begin{table}[ht]
\caption{The event horizon radii, photon sphere and shadow radii for a=2, c=1}
\centering
\begin{tabular}{c c c c c c c c c c  c}\\[1ex]
\hline
$b$ & 1.3 & 1.5 & 2 & 2.5 & 3 & 4 & 4.5 & 5 & 12  \\[0.5ex]
\hline
 $x_+$ & 1.05 & 1.30 & 1.78 & 2.20 & 2.62 & 3.43 & 3.83 & 4.23 & 9.76 \\[0.5ex]
\hline
 $x_p$ & 2.12 & 2.37 & 2.97 & 3.57 & 4.16 & 5.35 & 5.94& 6.53 & 14.78   \\[0.5ex]
\hline
 $x_s$ & 4.28 & 4.68 & 5.68 & 6.68 & 7.69 & 9.71 & 10.73 & 11.74 & 26.00 \\[0.5ex]
\hline
\end{tabular}
\end{table}
\begin{table}[ht]
\caption{The event horizon radii, photon sphere and shadow radii for b=c=1}
\centering
\begin{tabular}{c c c c c c c c c c  c}\\[1ex]
\hline
$a$ & 2.5 & 3 & 4.5 & 6.5 & 7 & 7.5 & 8.5 & 9 & 9.5 \\[0.5ex]
\hline
 $x_+$ & 1.34 & 1.74 & 2.67 & 3.77 & 4.04 & 4.31 & 4.83 & 5.09 & 5.35 \\[0.5ex]
\hline
 $x_p$ & 2.35 & 2.85 & 4.15 & 5.76 & 6.15 & 6.55 & 7.33 & 7.71 & 8.10 \\[0.5ex]
\hline
 $x_s$ & 4.54 & 5.31 & 7.44 & 10.16 & 10.83 & 11.50 & 12.83 & 13.49 & 14.50 \\[0.5ex]
\hline
\end{tabular}
\end{table}
According to Table 1, when the parameter $b$ is increased at fixed $a$ and $c$, the shadow radius $x_s$ increasing.
Tables 1 and 2 show that $x_s>x_+$ and the BH shadow radius is defined by the photon capture radius $r_s=x_s\sqrt[4]{\beta q_m^2}$. In accordance to Table 2 with increasing the parameter $a$ at fixed $b$ and $c$ the event horizon radii, the photon sphere radii and impact parameters are increased. When the parameters $a$ and $b$ are constants at variable $a$, the GB parameter $\alpha$ and nonlinearity parameter $\beta$ are fixed, increasing $a$ means increasing the BH mass $M$. Therefore, Table 2 shows that if the Schwarzschild BH mass $M$ is increased the event horizon radii, the photon sphere radii and impact parameters are increasing. It should be noted that to study the dependence of $r_+$, $r_p$ and $r_s$ on
$\alpha$ and $\beta$, one has to put different numerical values for $\alpha$ and $\beta$ in Eqs. (27) and (28).

It is worth noting that nonlinearity of the NED Lagrangian leads to the self-interaction, and photons do not propagate along null geodesics of metric (9) but they propagate along null geodesics of the effective metric \cite{Novello}, \cite{Novello1}. The event horizons are still determined by the zeros of the metric function of (9), $f(r_h)=0$. As pointed in \cite{Kocherlakota} the accuracy in the photon sphere radius and impact parameter within the current approach is not clear.

\section{The energy emission rate}

Let us study the particle emission rate of the BH depending on the model parameters.
The BH shadow is connected with the high energy absorption cross section for the observer located at infinity
\cite{Wei}–\cite{Belhaj}. At very high energies the absorption cross-section oscillates around an approximate value of the photon sphere $\sigma\approx \pi r_s^2$. The energy emission rate is given by
\begin{equation}
\frac{d^2E(\omega)}{dtd\omega}=\frac{2\pi^3\omega^3r_s^2}{\exp\left(\omega/T_H(r_+)\right)-1},
\label{31}
\end{equation}
where $\omega$ denotes the emission frequency and $T_H$ is the
Hawking temperature. Making use of the dimensionless variable $x=r/\sqrt[4]{\beta q_m^2}$ and Eq. (31)
one finds
\begin{equation}
\beta^{1/4}\sqrt{q_m}\frac{d^2E(\omega)}{dtd\omega}=
\frac{2\pi^3\varpi^3x_s^2}{\exp\left(\varpi/\bar{T}_H(x_+)\right)-1},
\label{32}
\end{equation}
where the Hawking temperature is given by Eq. (17), $\bar{T}_H(x_+)=\beta^{1/4}\sqrt{q_m}T_H(x_+)$ and $\varpi=\beta^{1/4}\sqrt{q_m}\omega$.
In Fig. 6, we plotted the radiation rate as a function of the dimensionless emission frequency $\varpi$  for $c=1$, $a=2$ and $b=5, 12, 17$.
\begin{figure}[h]
\includegraphics[height=4.0in,width=4.0in]{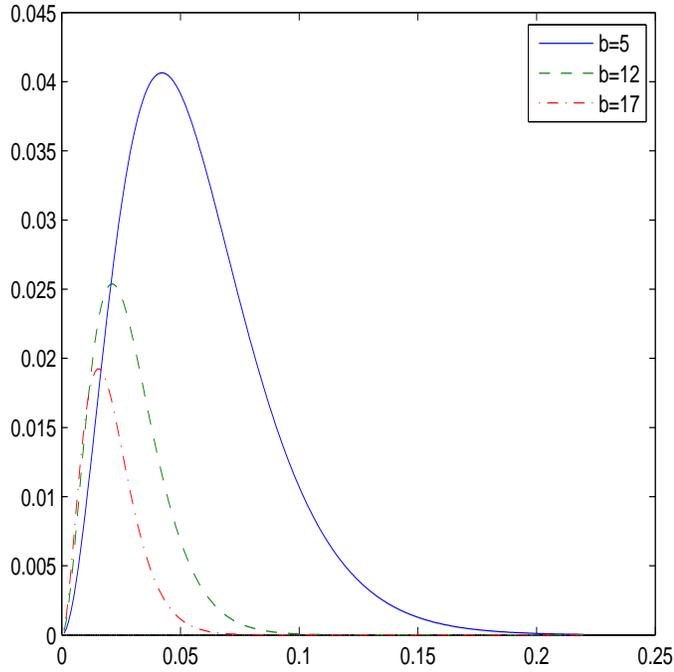}
\caption{\label{fig.6}The plot of the function $\beta^{1/4}\sqrt{q_m}\frac{d^2E(\omega)}{dtd\omega}$ vs. $\varpi$ for $b=5, 12, 17$, $a=2$, $c=1$.}
\end{figure}
Figure 6 shows that there is a peak of the energy emission rate for the BH. When the parameter $b$ increases, the peak of the energy emission rate decreases and goes to the low frequency. The BH has a bigger lifetime at a bigger parameter $b$. To investigate the dependence of the energy emission rate on parameters $\alpha$ and $\beta$, one should use numerical values for $\alpha$ and $\beta$.

\section{Quasinormal Modes}

Let us consider perturbations by a scalar massless field around the BH. The equation of motion for a scalar field is
\begin{equation}
\frac{1}{\sqrt{-g}}\partial_\mu\left(\sqrt{-g}g^{\mu\nu}\partial_\nu\Phi\right)=0,
\label{33}
\end{equation}
where the metric determinant for the spherically symmetric metric (9) is $g=-r^4\sin^2\vartheta$. To separate the variables we use the decomposition
\begin{equation}
\Phi=e^{-i\omega t}\frac{R(r)}{r}e^{im\phi} Y_{lm}(\vartheta),
\label{34}
\end{equation}
where $R(r)$ is the radial function and $Y_{lm}(\vartheta)$ are spherical harmonics. Introducing the tortoise coordinates $r_*=\int dr/f(r)$, the radial equation takes the form
\begin{equation}
\frac{d^2R(r_*)}{dr_*^2}+(\omega^2-V(r_*))R(r_*)=0,
\label{35}
\end{equation}
where the effective potential barrier is given by
\begin{equation}
V(r)=f(r)\left(\frac{f'(r)}{r}+\frac{l(l+1)}{r^2}\right),
\label{36}
\end{equation}
with $l$ being the multipole number $0,1,2...~$. In the terms of dimensionless variable $x=r/\sqrt[4]{\beta q_m^2}$, Eq. (36) reads
\begin{equation}
V(x)\sqrt{\beta}q_m=f(x)\left(\frac{f'(x)}{x}+\frac{l(l+1)}{x^2}\right).
\label{37}
\end{equation}
The effective potential is plotted in Fig. 7 for $a=b=2$, $c=1$ and $l=1,2,3$ and in Fig. 8 for $a=2$, $c=1$, $l=1$ and $b=3,4,5$.
\begin{figure}[h]
\includegraphics[height=4.0in,width=4.0in]{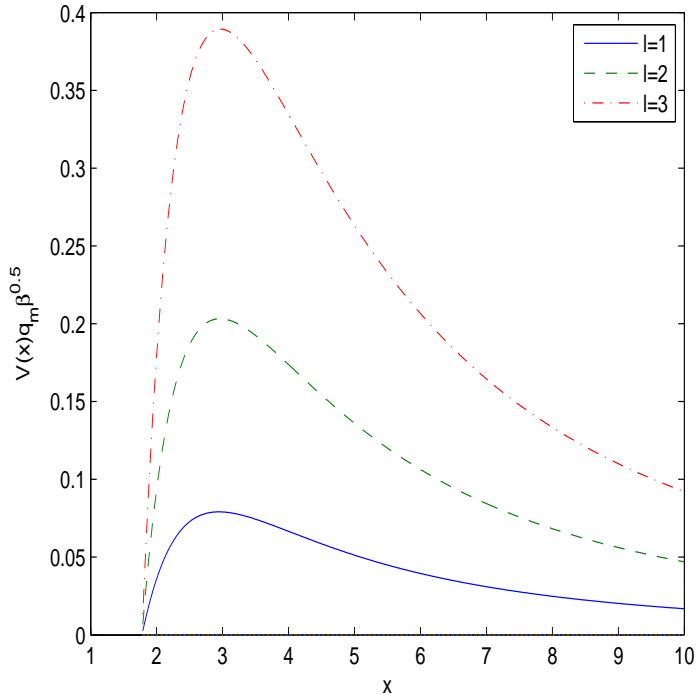}
\caption{\label{fig.7}The plot of the function $V(x)\sqrt \beta q_m$ for $a=b=2$, $c=1$, $l=1,2,3$.}
\end{figure}
\begin{figure}[h]
\includegraphics[height=4.0in,width=4.0in]{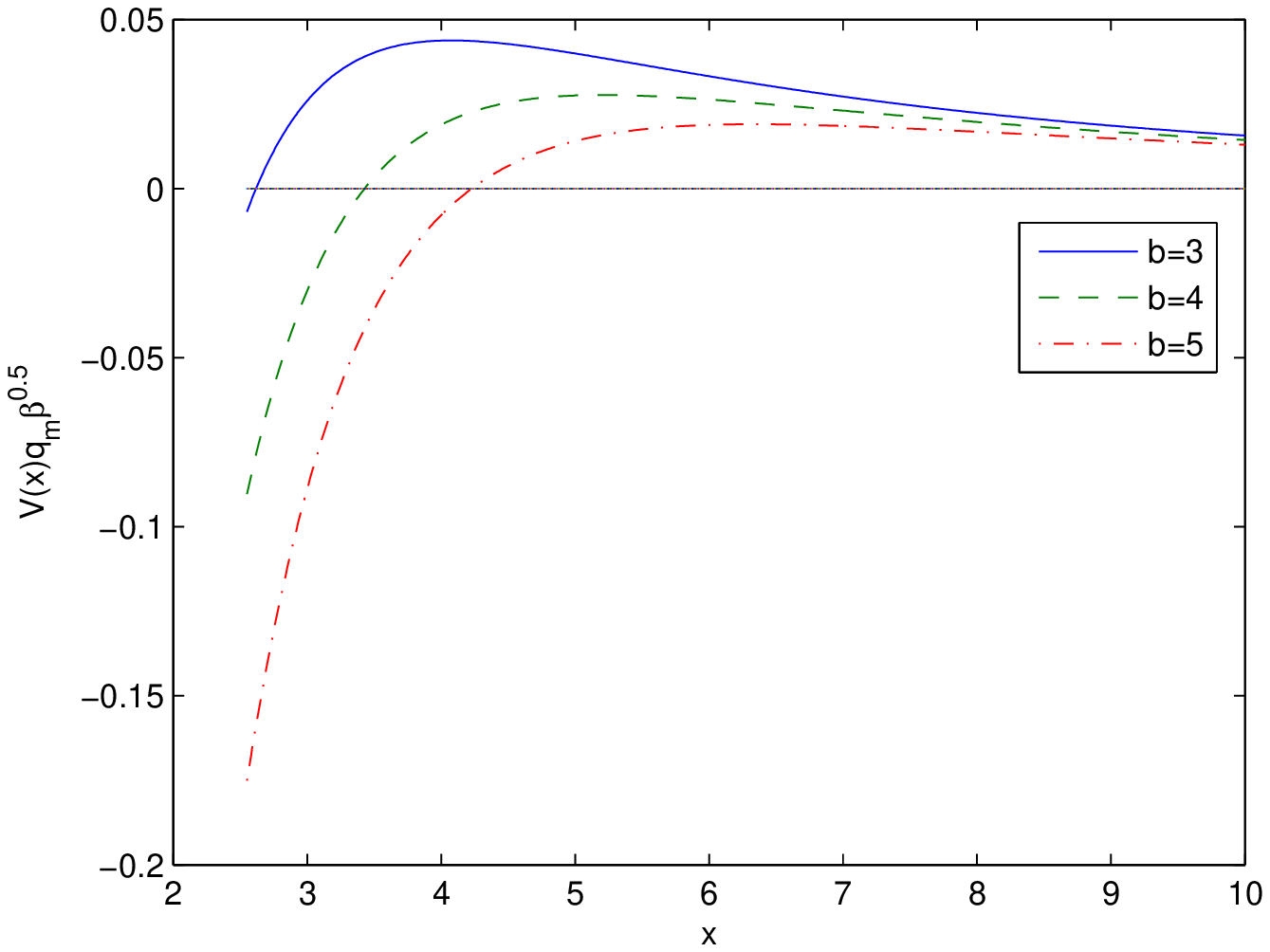}
\caption{\label{fig.8}The plot of the function $V(x)\sqrt\beta q_m$ for $a=2$, $c=1$, $l=1$, $b=3, 4, 5$.}
\end{figure}
Figure 7 shows that the effective potentials represent a potential barrier with a maximum. The height of the potential increases when
the $l$ increases. According to Fig. 8 height of the potential decreases if the parameter $b$ increases. It was shown that Re~$\omega$ in the eikonal limit is connected with the BH shadow radius \cite{Jusufi2}, \cite{Jusufi3}.
We will use the semi-analytical WKB approximation \cite{Iyer}, \cite{Kokkotas}. In the third order expansion the frequencies are \cite{Lobo} ($\omega=\omega_R-i\omega_I$, $\mbox{Im}~\omega=-\omega_I$)
\begin{equation}
\omega_R=\frac{1}{r_s}\left(l+\frac{1}{2}+{\cal O}(l^{-1})\right),
\label{38}
\end{equation}
\begin{equation}
\omega_I=\frac{2n+1}{2\sqrt{2}r_s}\sqrt{2f(r_p)-r_p^2f''(r_p)},
\label{39}
\end{equation}
where $n=0,1,2,...$ is the overtone number. In Table 3, one can find the real and imaginary parts of the frequencies versus the parameter $b$ at $a=2$, $c=1$, $n=0$, $l=10$.
\begin{table}[ht]
\caption{The real and the imaginary parts of the frequencies vs the parameter $b$ at $n=0$, $l=10$,  $a=2$, $c=1$}
\centering
\begin{tabular}{c c c c c c c c c c  c}\\[1ex]
\hline
$b$  & 1.5 & 2 & 2.5 & 3 & 4 & 5 & 12  \\[0.5ex]
\hline
 $\sqrt[4]{\beta q_m^2}\omega_R$  &2.24 & 1.85 & 1.57 & 1.37& 1.08 & 0.89 & 0.40 \\[0.5ex]
\hline
 $\sqrt[4]{\beta q_m^2}\omega_I$ & 0.15 & 0.14 & 0.12& 0.11 & 0.09 & 0.08 & 0.05 \\[0.5ex]
\hline
\end{tabular}
\end{table}
Because the imaginary parts of the frequencies ($\mbox{Im}~\omega=-\omega_I$) in Table 3 are negative the modes are stable and the real part represents the frequency of oscillations. Table 3 shows that when the parameter $b$ increases the real part of reduced frequencies  $\sqrt[4]{\beta q_m^2}\omega_R$ and $\sqrt[4]{\beta q_m^2}\omega_I$ decrease. In other words, decreasing the parameter $b$  the scalar perturbations oscillate with greater frequency and decay quickly.
As the model possesses many parameters $\alpha$, $\beta$, $M$, $q_m$ we do not study here the dependence of frequencies on these parameters in detail.

\section{Conclusion}

 The exact spherically symmetric and magnetically charged BH solution in 4D EGB gravity coupled to NED
proposed in \cite{Kruglov1} is  obtained.
The BH can have two, one or not horizons depending on the model parameters $\alpha$, $\beta$, $q_m$ and $M$.
It was demonstrated that the limit $r\rightarrow 0$ in the metric function results in the value $f(0)=1$ while in the pure 4D EGB theory this limit gives the non-physical value of $f(0)$. Thus, we have the regular BH  and the Reissner$-$Nordstr\"{o}m behaviour of the charged BH at infinity. We shown that WEC, DEC and SEC are satisfied.

We studied the thermodynamics and the thermal stability of magnetized BHs calculating the BH Hawking temperature and the heat capacity. It was shown that the second-order phase transitions occur in the points where the Hawking temperature possesses an extremum and the heat capacity has the discontinuity. It was demonstrated that the BHs are thermodynamically stable in some range of event horizon radii where the heat capacity and the Hawking temperature are positive. At the point where the Hawking temperature and heat capacity are zero we have the first-order phase transition and the BH remnant is created.
We obtained the logarithmic correction to the Bekenstein$-$Hawking entropy mimicking the quantum correction.
We show that when increasing the parameter $b$ the BH shadow radius is increased for some model parameters and the energy emission decreases.
It was demonstrated that the BH energy emission decreases with increasing the parameter $b$.
We shown that decreasing the parameter $b$  the scalar perturbations oscillate with greater frequency.

The attractive features of model considered here are as follows.

$\bullet$ At $\beta = 0$ or at $\beta {\cal F}<<1$ we recover the linear Maxwell
Lagrangian.

$\bullet$ The metric function obtained possesses a simple structure and is expressed
through elementary functions. It has regularity at $r\rightarrow 0$
($f(0)=1$). At $r\rightarrow \infty$ the Reissner$-$Nordstr\"{o}m behaviour
of the charged BH is recovered.

$\bullet$ The WEC and DEC are satisfied.

$\bullet$  The logarithmic correction to the Bekenstein-Hawking entropy obtained
mimics the quantum corrections.

$\bullet$ The BHs are locally stable at some range of event horizon radii.

$\bullet$ In this model the first-order phase transition occurs and
the BH remnant is created which can be a candidate for dark matter.

It is worth noting that the solution for the metric function in the 4D EGB gravity coupled to Born$-$Infeld NED \cite{Wei1} is more complicated compared to our model and contains the special hypergeometric function. In \cite{Jusufi}, \cite{Ghosh} solutions for metric functions in 4D EGB gravity with the exponential NED have simple interesting structures.

If 4D EGB theory suggested in Ref. 2 is unjustified, we can postulate the metric function (10). The metrics in \cite{Bardeen}, \cite{Hayward}, \cite{Frolov} also were postulated without obtaining the source of gravity. Our metric function (10) is new, and of interest to study BH properties.

\vspace{3mm}
\textbf{Appendix: The energy conditions}
\vspace{3mm}

The symmetrical energy-momentum tensor (6) with the spherically symmetry results to $T_t^{~t}=T_r^{~r}$ and the radial pressure gives $p_r=-T_r^{~r}=-\rho$. The tangential pressure is given by $p_\perp=-T_\vartheta^{~\vartheta}=-T_\phi^{~\phi}$ so that \cite{Dymnikova}
\begin{equation}
p_\perp=-\rho-\frac{r}{2}\rho'(r),
\label{40}
\end{equation}
where the prime denotes the derivative with respect to $r$. The Weak Energy Condition (WEC) is satisfied if and only if $\rho\geq 0$ and $\rho+p_k\geq 0$ (k=1,2,3) \cite{Hawking}. This guarantees that the energy density is positive. According to Eq. (7) $\rho\geq 0$. From Eq. (7) we obtain
\begin{equation}
\rho'(r)=-\frac{q_m^2(2r^2+q_m\sqrt{\beta})}{r^3(r^2+q_m\sqrt{\beta})^2}\leq 0.
\label{41}
\end{equation}
Thus, we have $\rho\geq 0$, $\rho+p_r\geq 0$, $\rho+p_\perp\geq 0$ and WEC is satisfied for our model. The Dominant Energy Condition (DEC) holds when \cite{Hawking} $\rho\geq o$, $\rho+p_k\geq 0$, $\rho-p_k\geq 0$. Because these conditions include WEC, we should verify the condition $\rho-p_\perp\geq 0$. From Eqs. (7), (40) and (41) one finds
\begin{equation}
\rho-p_\perp=\frac{q_m^3\sqrt{\beta}}{2r^2(r^2+q_m\sqrt{\beta})^2}\geq 0.
\label{42}
\end{equation}
As a result, DEC is satisfied. Therefore, the sound speed is less than the light speed.
The Strong Energy Condition (SEC) requires the condition $\rho+\sum_{k=1}^3 p_k\geq 0$ \cite{Hawking}. With the help of Eqs. (7), (40) and (41) we obtain
\begin{equation}
\rho+\sum_{k=1}^3 p_k=2p_\perp=\frac{q_m^2}{(r^2+q_m\sqrt{\beta})^2}\geq 0.
\label{43}
\end{equation}
As a result, SEC is satisfied. SEC defines the  gravity acceleration.


\begin{thebibliography}{99}

\bibitem{Witten} D. J. Gross and E. Witten, Nucl. Phys. B \textbf{277}, 1 (1986); D. J. Gross and J. H. Sloan, Nucl. Phys. B \textbf{291}, 41 (1987); R. R. Metsaev and A. A. Tseytlin, Phys. Lett. B \textbf{191}, 354 (1987); B. Zwiebach, Phys. Lett. B \textbf{156}, 315 (1985); R. R. Metsaev and A. A. Tseytlin, Nucl. Phys. B \textbf{293}, 385 (1987).
\bibitem{Glavan} D. Glavan and C. Lin, Phys. Rev. Lett. \textbf{124}, 081301 (2020).
\bibitem{Deser}D. G. Boulware and S. Deser, Phys. Rev. Lett. \textbf{55}, 2656 (1985); J. T. Wheeler, Nucl. Phys. B \textbf{268}, 737 (1986);  R.C. Myers and J.Z. Simon, Phys. Rev. D \textbf{38}, 2434 (1988).
\bibitem{Lovelock}D. Lovelock, J. Math. Phys. \textbf{12}, 498 (1971).
\bibitem{Cai}R. G. Cai, L. M. Cao, and N. Ohta, JHEP \textbf{1004}, 082 (2010).
\bibitem{Cai1} R.-G. Cai, Phys. Lett. B \textbf{733}, 183 (2014).
\bibitem{Cognola} G. Cognola, R. Myrzakulov, L. Sebastiani, and S. Zerbini, Phys. Rev. D \textbf{88}, 024006   (2013),
\bibitem{Fernandes} P. G. S. Fernandes, Phys. Lett. B \textbf{805} 135468 (2020).
\bibitem{Jusufi}K. Jusufi, Ann. Phys. \textbf{421}, 168285 (2020).
\bibitem{Ghosh} S. G. Ghosh, D. V. Singh, R. Kumar, and S. D. Maharaj, Ann. Phys. \textbf{424}, 168347 (2021).
\bibitem{Ghosh1} S. G. Ghosh and S. D. Maharaj, Phys. Dark Univ. \textbf{30}, 100687 (2020).
\bibitem{Ghosh2} R. Kumar and S. G. Ghosh, JCAP \textbf{07}, 053 (2020).
\bibitem{Jin} X. H. Jin, Y. X. Gao, and D. J. Liu, Int. J. Mod. Phys. D \textbf{29}, 2050065 (2020).
\bibitem{Jusufi1} K. Jusufi, A. Banerjee, and S. G. Ghosh, Eur. Phys. J. C \textbf{80}, 698 (2020).
\bibitem{Guo} M. Guo and P. Li, Eur. Phys. J. C \textbf{80}, 588 (2020).
\bibitem{Zhang} C. Zhang, S. Zhang, P. Li, and M. Guo, JHEP \textbf{08}, 105 (2020).

\bibitem{Odintsov} S. Odintsov, V. Oikonomou, and F. Fronimos, Nucl. Phys. B \textbf{958}, 115135 (2020).
\bibitem{Ai} W. Ai, Commun. Theor. Phys. \textbf{72}, 095402 (2020).
\bibitem{Fernandes1} P. G. Fernandes, P. Carrilho, T. Clifton, and D. J. Mulryne, Phys. Rev. D \textbf{102}, 024025 (2020).
\bibitem{Hennigar} R. A. Hennigar, D. Kubiznak, R. B. Mann, and C. Pollack, JHEP \textbf{2020}, 27 (2020).
\bibitem{Martinez}H. A. Gonzalez, M. Hassaine, and C. Martinez, Phys. Rev. D \textbf{80}, 104008 (2009).
\bibitem{Olea}O. Miskovic and R. Olea, Phys. Rev. D \textbf{83}, 024011 (2011).
\bibitem{Hendi2}S. H. Hendi, S. Panahiyan, and M. Momennia, Int. J. Mod. Phys. D \textbf{25}, 1650063 (2016).
\bibitem{Garcia}D. Rubiera-Garcia, Phys .Rev. D \textbf{91}, 064065 (2015).
\bibitem{Hendi}S. H. Hendi, B. Eslam, and S. Panahiyan, Fortsch. Phys. \textbf{66}, 1800005 (2018).
\bibitem{Odintsov1}S. Nojiri and S. D. Odintsov, Phys. Rev. D \textbf{96}, 104008 (2017).
\bibitem{Nam1}C. H. Nam, Gen. Rel. Grav. \textbf{51}, 104 (2019).
\bibitem{Nam}S. Hyun and C. H. Nam, Eur. Phys. J. C \textbf{79}, 737 (2019).
\bibitem{Stuchlik}M. S. Churilova and Z. Stuchlik, Ann. Phys. \textbf{418}, 168181 (2020).
\bibitem{Lobo}K. Jafarzade, M. K. Zangeneh, and F. S. N. Lobo, Optical features of AdS black holes in the novel 4D Einstein$-$Gauss$-$Bonnet gravity coupled to nonlinear electrodynamics, arXiv:2009.12988 [gr-qc].
\bibitem{Tomozawa} Y. Tomozawa, Quantum corrections to gravity, arXiv:1107.1424 [gr-qc].
\bibitem{Tekin}M. Gurses, T. C. Sisman, and B. Tekin, Phys. Rev. Lett. \textbf{125},  149001 (2020).
\bibitem{Tekin1} M. Gurses, T. C. Sisman, and B. Tekin, Eur. Phys. J. C \textbf{80}, 647 (2020).
\bibitem{Mahapatra} S. Mahapatra, Eur. Phys. J. C \textbf{80}, 992 (2020).
\bibitem{Tian} S. X. Tian, Z.-H. Zhu, Non-full equivalence of the four-dimensional Einstein-Gauss-Bonnet gravity and Horndeksi gravity for Bianchi type I metric, arXiv:2004.09954 [gr-qc].
\bibitem{Arrechea} J. Arrechea, A. Delhom, and A. Jiménez-Cano, Chin. Phys. C \textbf{45}, 013107 (2021).
\bibitem{Arrechea}J. Arrechea, A. Delhom, and A. Jiménez-Cano, Phys. Rev. Lett. \textbf{125}, 149002 (2020).
\bibitem{Hohmann} M, Hohmann and C. Pfeifer,  Eur. Phys. J. Plus \textbf{136}, 180 (2021).
\bibitem{Aoki} K. Aoki, M. A. Gorji, and S. Mukohyama, Phys. Lett. B \textbf{810}, 135843 (2020).
\bibitem{Aoki1} K. Aoki, M. A. Gorji, and S. Mukohyama, JCAP \textbf{2009}, 014 (2020).
\bibitem{Wei1} K. Yang, B. M. Gu, S .W. Wei, and Y. X. Liu, Eur. Phys. J. C  \textbf{80}, 662 (2020).
\bibitem{Lobo1} K. Jafarzade, M. K. Zangeneh, F. S. N. Lobo, Shadow, deflection angle and quasinormal modes of Born-Infeld charged black holes, arXiv:2010.05755 [gr-qc].
\bibitem{Schutz}B. F. Schutz and C. M. Will, Astrophys. J. Lett. \textbf{291}, L33 (1985).
\bibitem{Iyer}S. Iyer and C. M. Will, Phys. Rev. D \textbf{35}, 3621 (1987).
\bibitem{Konoplya}R. A. Konoplya, A. F. Zinhailo, and Z. Stuchlik, Phys. Rev. D \textbf{102}, 044023 (2020).
\bibitem{Kruglov1}S. I. Kruglov, Ann. Phys. (Berlin) \textbf{529}, 1700073 (2017).
\bibitem{Kruglov} S. I. Kruglov, Phys. Rev. D \textbf{94}, 044026 (2016); ibid \textbf{92}, 123523 (2015);
 Ann. Phys. (Berlin) \textbf{528}, 588 (2016);  Int. J. Mod. Phys. D \textbf{26}, 1750075 (2017);
Ann. Phys. \textbf{353}, 299 (2015); ibid \textbf{409}, 167937 (2019);
 Eur. Phys. J. C \textbf{80}, 250 (2020); Grav. Cosmol. \textbf{25}, 190 (2019); Gen. Rel. Grav. \textbf{51}, 121 (2019).
\bibitem{Habib} S. H. Mazharimousavi and M. Halilsoy, Phys. Lett.B \textbf{796}, 123 (2019).
\bibitem{Bronnikov} K. A. Bronnikov, Phys. Rev. D \textbf{63}, 044005 (2001).
\bibitem{Bronnikov1} K. A. Bronnikov, Phys. Rev. D \textbf{101}, 128501 (2020); Phys. Rev. Lett. \textbf{85}, 4641 (2000); Int. J. Mod. Phys. D \textbf{27}, 1841005 (2018); Grav. Cosmol. \textbf{23}, 343 (2017).
\bibitem{Bronnikov2}   K. A. Bronnikov, G. N. Shikin, and E. N. Sibileva, Grav. Cosmol. \textbf{9}, 169 (2003).
\bibitem{Burinskii} A. Burinskii and S. R. Hildebrandt, Phys. Rev. D \textbf{65}, 104017 (2002).
\bibitem{Diaz} J. Diaz-Alonso and D. Rubiera-Garcia, Phys. Rev. D \textbf{81}, 064021 (2010).
\bibitem{Novello}    M. Novello, V. A. De Lorenci, J. M. Salim, and R. Klippert, Phys. Rev. D \textbf{61}, 045001 (2000).
\bibitem{Novello1} M. Novello, S. E. Perez Bergliaffa, and J. M. Salim, Class. Quant. Grav. \textbf{17}, 3821 (2000).
\bibitem{Quiros} R. Garcia-Salcedo, T. Gonzalez, and I. Quiros, Phys. Rev. D \textbf{89}, 084047 (2014).
\bibitem{Odintsov3} M. Cvetic, S. Nojiri, and S. D. Odintsov, Nucl. Phys. B \textbf{628}, 295 (2002).
\bibitem{Event1} Event Horizon Telescope collaboration, K. Akiyama et al., Astrophys. J. \textbf{875}, L5 (2019).
\bibitem{Synge} J. L. Synge, Mon. Not. Roy. Astron. Soc. \textbf{131}, 463 (1966).
\bibitem{Zhang1} M. Zhang and M. Guo, Eur. Phys. J. C \textbf{80},  790 (2020).
\bibitem{Kocherlakota}P. Kocherlakota and L. Rezzolla, Phys. Rev. D, \textbf{6}, 064058 (2020).
\bibitem{Wei} S. W. Wei and Y. X. Liu, JCAP 11, 063 (2013).
\bibitem{Belhaj1} A. Belhaj, M. Benali, A. El Balali, H. El Moumni, and
S. E. Ennadifi, Class. Quant. Grav. \textbf{37}, 215004 (2020).
\bibitem{Belhaj} A. Belhaj, M. Benali, A. El Balali, W. El Hadri, and H. El
Moumni, arXiv:2007.09058.
\bibitem{Jusufi2} K. Jusufi, Phys. Rev. D \textbf{101}, 084055 (2020).
\bibitem{Jusufi3} K. Jusufi, Phys. Rev. D \textbf{101}, 124063 (2020).
\bibitem{Kokkotas}K. Kokkotas and B. F. Schutz, Phys. Rev. D \textbf{37}, 3378 (1988).
\bibitem{Bardeen} J. Bardeen, Proc. GR5, Tiflis, USSR (1968), p.174.
\bibitem{Hayward} S. A. Hayward, Phys. Rev. Lett. \textbf{96}, 031103 (2006).
\bibitem{Frolov} V. P. Frolov, Phys. Rev. D \textbf{94}, 104056 (2016).
\bibitem{Dymnikova} I. Dymnikova, Class. Quant. Grav. \textbf{21}, 4417 (2004).
\bibitem{Hawking} S. W. Hawking and G. F. R. Ellis, \textit{The large scale structure of space-time}, Canbridge Univ. Press (1973).

\end{thebibliography}
\end{document}